\documentclass[showpacs,preprintnumbers,showkeys]{revtex4-1}
\usepackage{graphicx}
\usepackage{bm}
\usepackage{amsmath}
\usepackage[bookmarks=false]{hyperref}

\begin{document}

\title{Demonstration of Deutsch's Algorithm on a Stable Linear-Optical Quantum Computer}
\author{Pei Zhang,$^{1} $\footnote[1]{zhang.pei@stu.xjtu.edu.cn}
Rui-Feng Liu,$^{1} $ Yun-Feng Huang,$^{2} $ Hong Gao,$^{1} $ and Fu-Li Li$^{1} $}

\address{$^{1}$MOE Key Laboratory for Nonequilibrium Synthesis and
Modulation of Condensed Matter, Department of Applied Physics, Xi'an
Jiaotong University, Xi'an 710049, China\\
$^{2}$Key Laboratory of Quantum Information, University of Science and
Technology of China, CAS, Hefei 230026, China}

\begin{abstract}
We report an experimental demonstration of quantum Deutsch's algorithm
using a linear-optical system. By employing photon polarization and spatial
modes, we implement all balanced and constant functions for a quantum computer.
The experimental system is very stable, and the experimental data are
in excellent accordance with the theoretical results.
\end{abstract}

 \maketitle

Quantum computation may solve some complex computational problems
and hit the security of the classical cryptography. It has attracted
much interest to investigate quantum algorithms and to realize
quantum hardware, which are very important to quantum information
processing and quantum computation. The first quantum algorithm
was proposed by Deutsch in 1985 \cite{Deutsch85}, then extended
by Deutsch and Jozsa in 1992 \cite{Deutsch92}.
The realizations of quantum Deutsch's algorithm on quantum computers
have been examined in many physical systems,
including ion traps \cite{Cirac95}, nuclear spins in magnetic
resonance \cite{Gersh97}, super conducting resonators
\cite{akamura99}, semiconductor quantum dots \cite{ayashi03},
neutral atoms \cite{Jaksch04}, and linear optics \cite{Mohseni,Oliveira05,Tame07}. In linear
optical systems, it is easy to deal with entanglement and
decoherence, and the incorporation of detection and post selection
make it possible to achieve all-optical quantum computers \cite{Knill01},
so linear optical systems are a good candidate for
implementing quantum algorithms \cite{Ahn00, Bhatt02}.
The system of single-photon few-qubit has been used to build the
deterministic quantum information processor (QIP), and few-qubit
QIPs have drawn much attention for application in quantum
optics and quantum computation \cite{Mitsumori,Chen,Walborn,Genovese,Hogg,Kim,PRL93070502}.
Oliveira \emph{et al.} experimentally tested Deutsch's algorithm using a
single-photon two-qubit (SPTQ) system in 2005 \cite{Oliveira05}. However,
the light source in their experiment was a bright coherent light instead of a single photon,
and they realized the four relevant operations using a phase-sensitive
Mach-Zehnder interferometer, which needs additional phase stabilization.
In this Brief Report, we experimentally demonstrate Deutsch's algorithm
using a more robust setup \cite{PRL93070502} at the single-photon level.
By employing photon polarization and spatial modes as a SPTQ system,
we implement all balanced and constant functions for quantum computer.
The experimental system is very stable, and the experimental data are
excellent in accordance with the theoretical results. Furthermore,
we also introduce a phase variation in the input spatial qubit,
which helps us easily to study the differences of all input states for the algorithm.

\begin{figure}[tbh]
\includegraphics[width=6cm]{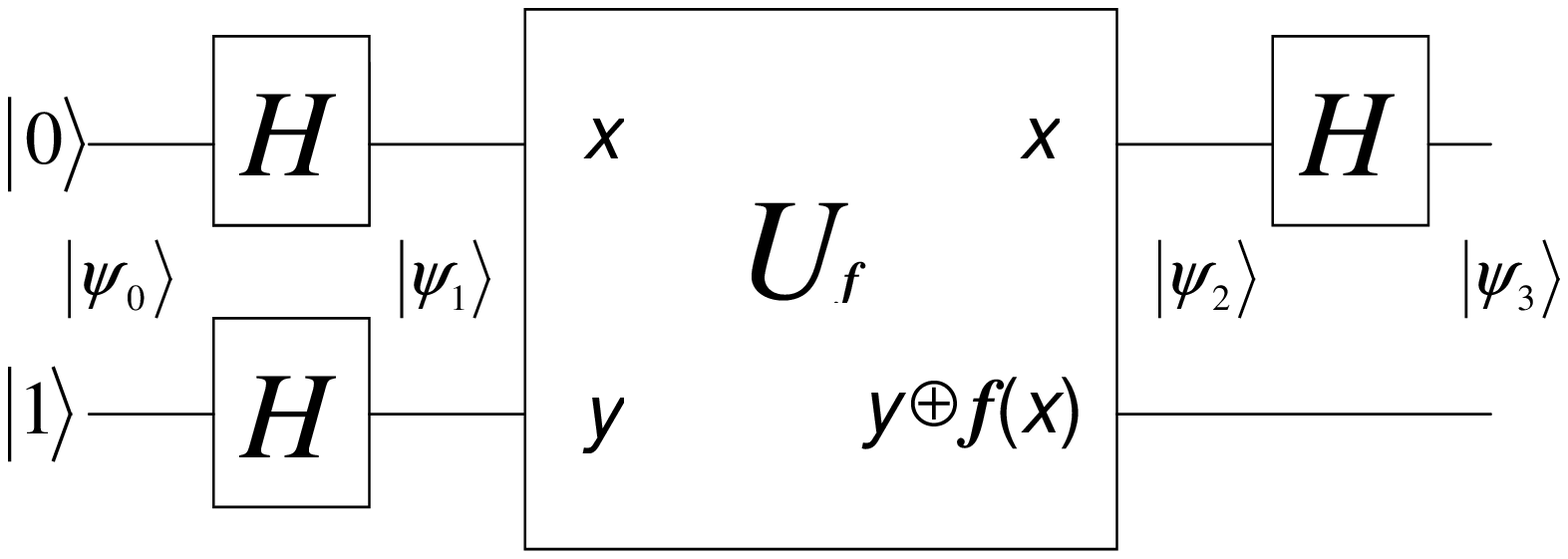}
\caption{Quantum circuit of Deutsch's algorithm. $H$ is the
Hadamard gate. }
\end{figure}

Deutsch's algorithm combines quantum parallelism with a
property of quantum interference. Suppose we are given a boolean function
$f(x)$, where $x$ is either 0 or 1. What we want to know is whether $f(x)$ is a constant function or a balanced function, where constant
function means $f(x)=0$ or $f(x)=1$ [or $f(0)=f(1)$] and balanced
function means $f(x)=x$ or $f(x)=inv(x)$ [or $f(0)\neq f(1)$] and
`$inv$' is the inversion operation. The classic computer has to run
the $f(x)$ twice to distinguish a balanced function from a constant
function, while a quantum computer does the job in just one go.
Fig. 1 is the quantum circuit implementing Deutsch's algorithm \cite{Nielsen00}%
. $U_{f}$ is the quantum operation which takes inputs $\left\vert
x,y\right\rangle $ to $\left\vert x,y\oplus f(x)\right\rangle $. A brief explanation is given subsequently.
The initial state is $\left\vert \Psi _{0}\right\rangle =\left\vert 0\right\rangle \left\vert
1\right\rangle $. After the Hadamard transformation (\emph{H}), we get $\left\vert \Psi _{1}\right\rangle =(\left\vert 0\right\rangle
+\left\vert 1\right\rangle )(\left\vert 0\right\rangle -\left\vert
1\right\rangle )/2.$
Applying $U_{f}$ to $\left\vert \Psi _{1}\right\rangle $, we obtain
$\left\vert \Psi _{2}\right\rangle$ to be one of two possible states, depending on $f(x)$:
\begin{equation}
\makeatletter
\let\@@@alph\@alph
\def\@alph#1{\ifcase#1\or \or $'$\or $''$\fi}\makeatother
\left\vert \Psi _{2}\right\rangle=
\begin{cases}
\pm ( \left\vert
0\right\rangle +\left\vert 1\right\rangle ) (
\left\vert 0\right\rangle -\left\vert 1\right\rangle )/2,& f(0)=f(1), \\
\pm ( \left\vert
0\right\rangle -\left\vert 1\right\rangle ) (
\left\vert 0\right\rangle -\left\vert 1\right\rangle )/2,
&f(0)\neq f(1).
\end{cases}
\makeatletter\let\@alph\@@@alph\makeatother
\end{equation}
The final Hadamard gate is applied on the first qubit,%
\begin{equation}
\makeatletter
\let\@@@alph\@alph
\def\@alph#1{\ifcase#1\or \or $'$\or $''$\fi}\makeatother
\left\vert \Psi _{3}\right\rangle=
\begin{cases}
\pm \left\vert 0\right\rangle %
( \left\vert 0\right\rangle -\left\vert 1\right\rangle )
/\sqrt{2},&f(0)=f(1), \\
\pm \left\vert 1\right\rangle (
\left\vert 0\right\rangle -\left\vert 1\right\rangle )/\sqrt{2}
, &f(0)\neq f(1),
\end{cases}
\makeatletter\let\@alph\@@@alph\makeatother
\end{equation}
so we can determine $f(x)$ to be balanced or constant by only measuring the
first qubit once.

From the preceding description, to physically test the algorithm, we need a device which
can implement the $U_{f}$ operations for the four possible functions.
All the possible $f(x)$ functions and $U_{f}$ operations are summarized in Table I.
In the first case of $U_{f}=I$, it means that the second qubit never changes, whether the first qubit is 0 or 1,
so this can be recognized as an identity operation to the two qubits.
The second case shows that $U_{f}$ is a NOT gate. The second qubit always flips, no matter what the first qubit is.
In the third case, $U_{f}$ is a controlled-NOT (CNOT) gate. The second qubit flips when the first
qubit is 1. In the last case, $U_{f}$ is a zero-controlled-NOT (Z-CNOT) gate, where the second qubit flips when the first
qubit is 0. For these four different $U_{f}$ operations, identity operation and NOT operation are very simple to be realized, and the Z-CNOT
gate can be obtained from a CNOT gate with some small changes. So the CNOT gate is the fundamental and essential part to execute
Deutsch's algorithm. In this context, we start with a
CNOT gate realized by employing polarization and spatial positions of photons
\cite{PRL93070502}, construct the four different gates and $U_{f}$ operations, and carry out Deutsch's algorithm.
\begin{table}
\caption{\label{tab:journals}Four different cases of Deutsch's algorithm}
\begin{ruledtabular}
\begin{tabular}{llll}
\textbf{Class}    &\textbf{Function} &\textbf{Operation} &$\boldsymbol{U_f}$\\
\itshape Constant &$f(x)=0$       &$\left\vert x,y\right\rangle\rightarrow \left\vert x,y\oplus 0\right\rangle$           & \emph{I (Identity)}     \\
\itshape Constant  &$f(x)=1$      &$\left\vert x,y\right\rangle\rightarrow \left\vert x,y\oplus 1\right\rangle$           & NOT     \\
\itshape Balanced  &$f(x)=x$      &$\left\vert x,y\right\rangle\rightarrow \left\vert x,y\oplus x\right\rangle$           & CNOT     \\
\itshape Balanced  &$f(x)=inv(x)$ &$\left\vert x,y\right\rangle\rightarrow \left\vert x,y\oplus (x\oplus1)\right\rangle$  & Z-CNOT         \\
\end{tabular}
\end{ruledtabular}
\end{table}

The CNOT gate is shown in Fig. 2. The dove prism (DP) is
inclined at a $45^{\circ}$
angle relative to the horizontal plane [shown in Fig. 2(a)], so the images
which pass through it from left to right will be rotated by $90%
^{\circ }$. Suppose the polarized beam splitter (PBS) here transmits horizontal-polarized ($H$)
photons and reflects vertical-polarized ($V$) ones.
So the $H$ photons travel counterclockwise, while
the $V$ photons travel clockwise. With a DP
inclined at $45^{\circ }$, the spatial mode of $H$
($V$) photons is oriented $90^{\circ }$ ($-90^{\circ }$).
Specifically, the left-right ($l$-$r$) section of the input photons is
rotated into the down-up ($d$-$u$) section of the output beam for
$H$ photons but into the $u$-$d$ section for $V$
photons [shown in Fig. 2(b)]. If we define photon polarization as the control qubit ($V\rightarrow0$ and $H\rightarrow1$)
and spatial mode as the target qubit ($l\&u\rightarrow0$ and $r\&d\rightarrow1$),
the CNOT operation can be described as follow:
\begin{equation}
\left\vert V\right\rangle \left\vert l\right\rangle \rightarrow
\left\vert V\right\rangle \left\vert u\right\rangle ,\text{ \ \ \ \ \ }%
\left\vert V\right\rangle \left\vert r\right\rangle \rightarrow
\left\vert V\right\rangle \left\vert d\right\rangle ,\text{ \ \ \ \ \ }%
\left\vert H\right\rangle \left\vert l\right\rangle \rightarrow
\left\vert H\right\rangle \left\vert d\right\rangle ,\text{ \ \ \ \ \ }%
\left\vert H\right\rangle \left\vert r\right\rangle \rightarrow
\left\vert H\right\rangle \left\vert u\right\rangle .
\end{equation}%
For the Z-CNOT gate, we should realize the following transition:
\begin{equation}
\left\vert V\right\rangle \left\vert l\right\rangle \rightarrow
\left\vert V\right\rangle \left\vert d\right\rangle ,\text{ \ \ \ \ \ }%
\left\vert V\right\rangle \left\vert r\right\rangle \rightarrow
\left\vert V\right\rangle \left\vert u\right\rangle ,\text{ \ \ \ \ \ }%
\left\vert H\right\rangle \left\vert l\right\rangle \rightarrow
\left\vert H\right\rangle \left\vert u\right\rangle ,\text{ \ \ \ \ \ }%
\left\vert H\right\rangle \left\vert r\right\rangle \rightarrow
\left\vert H\right\rangle \left\vert d\right\rangle .
\end{equation}%
Similar to the implementation of the CNOT gate, if we set DP at $-45^{\circ }$, the spatial mode of $H$ ($V$) polarized
photons will be oriented $-90^{\circ }$ ($90^{\circ }$), and it will be a Z-CNOT gate.
The CNOT (Z-CNOT) gate is a
polarization Sagnac interferometer in our setup, and the two counter-propagating
photons always undergo the same amount of phase disturbance. So this optical
CNOT (Z-CNOT) gate has an inherent stability which requires no active stabilization.
\begin{figure}[tbh]
\includegraphics[width=6cm]{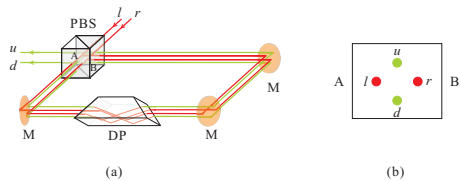}
\caption{(Color online) (a) Optical implementation of CNOT gate by
employing polarization and spatial positions of photons. The dove prism
(DP) is inclined at 45$^{\circ }$. Red lines (heavy gray) show the left ($l$) and right ($r$) spatial modes,
and green lines (light gray) show the up ($u$) and down ($d$) spatial modes. (b) Spatial positions
of input and output beams on the splitting plane AB of the polarized beam splitter (PBS).}
\end{figure}

We experimentally realize Deutsch's algorithm using the CNOT gate mentioned earlier.
The experimental setup is shown in Fig. 3.
Our source is a He-Ne laser (MELLES GRIOT, 05-LHP-171) with deep attenuation to
about 150,000 photon counts per second, which means that the mean distance between two photons
is about 2000 m (much bigger than our experimental setup length 0.5 m),
and the two-photon probability is $2.5\times10^{-4}$.
All the PBS are
quasi symmetric and transmit $H$ photons while reflecting
$V$ photons. A polarizer and half wave-plate (HWP$_1$) are used to prepare photon polarization states.
Here we prepare the initial polarization of photons as $V$.
A $50\%$ beam splitter (BS) and a mirror (M) are used to prepare the photon
spatial-mode states. The piezo-transmitter (PZT) on the first mirror is used
to control the relative phase $\varphi$ between two spatial modes. HWP$_2$ and HWP$_3$ at $22.5^{\circ }$ are
used as the polarization Hadamard gates.
The state after HWP$_2$ can be written as:
\begin{equation}
\left\vert \psi _{1}\right\rangle =(\left\vert
V\right\rangle +\left\vert H\right\rangle ) (\left\vert
l\right\rangle +e^{i\varphi}\left\vert r\right\rangle )/2;
\end{equation}%
in particular when $\varphi=\pi$, $\left\vert \psi _{1}\right\rangle$ is equal to
$\left\vert \Psi _{1}\right\rangle$, which is mentioned erlier. So this single-photon two-qubit state
can be used as the input state of Deutsch's algorithm as we described in Fig. 1.
Then this state will be evolved by the $U_f$ operation.
The detection part consists of a Hadamard gate (HWP$_3$), PBS$_2$, and two single-photon detectors (D$_1$ and D$_2$),
which detect the photon's polarization state (the first qubit of the output state).
The key point to carry out Deutsch's algorithm is how to realize the four different cases of $U_f$ operation.
We will discuss these four $U_f$ operations subsequently.
\begin{figure}[tbh]
\includegraphics[width=6cm]{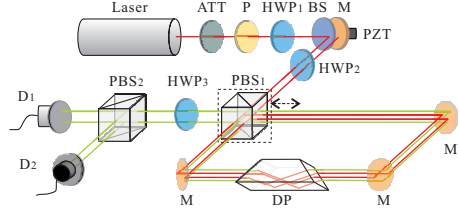}
\caption{(Color online) Experimental setup of Deutsch's algorithm. The source
is a He-Ne laser (MELLES GRIOT, 05-LHP-171) of wavelength 632.8 nm.
We attenuate the coherent light into a single-photon level by using some neutral attenuators
denoted by ATT. P denotes the polarizer for initial state preparation. Three half wave-plates (HWP) and two polarized beam splitters
(PBS) are used in the setup. A PZT actuator is used to modulate the
phase $\varphi$ between $l$ and $r$ paths. The dove plate (DP) is set at $45^{\circ }$.
Detectors (D$_1$ and D$_2$) are
single-photon counting modules (SPCM-AQRH-14-FC). All the mirrors are marked as M.}
\end{figure}

In the constant-function case, $U_f$ can be an identity or NOT operation.
For an identity operation, we can simply remove PBS$_1$ in our setup and set DP at $-45^{\circ }$. Therefore, photons in $l$ or $r$
will always undergo a counter clockwise route and be output in $u$ or $d$, respectively, without the effect of polarization.
This means that the target qubit (spatial mode of photons) will not change with control qubit (polarization of photons).
We can deduce the process as follow:
\begin{eqnarray}
&&\left\vert \psi _{1}\right\rangle \overset{I}{\longrightarrow }%
\left\vert \psi _{2}\right\rangle=(\left\vert
V\right\rangle +\left\vert H\right\rangle) (\left\vert
u\right\rangle +e^{i\varphi}\left\vert d\right\rangle )/2 \notag\\
&&\text{ \ \ \ \ }\overset{HWP_3}{\longrightarrow }\left\vert \psi _{3}\right\rangle= \left\vert
V\right\rangle (\left\vert
u\right\rangle +e^{i\varphi}\left\vert d\right\rangle) / \sqrt{2},
\end{eqnarray}%
whereas for the NOT operation, we can remove PBS$_1$ in our setup and set DP at $45^{\circ }$.
Then $\left\vert l \right\rangle$ is converted to $\left\vert u \right\rangle$
and $\left\vert r\right\rangle$ is converted to $\left\vert d \right\rangle$.
Applying a Hadamard gate (HWP3), we can obtain
\begin{eqnarray}
\left\vert \psi _{3}\right\rangle= \left\vert
V\right\rangle(\left\vert
d\right\rangle +e^{i\varphi}\left\vert u\right\rangle) /\sqrt{2}.
\end{eqnarray}%
For the preceding two cases, we can only detect the polarization qubits;
the results are same without any changes when we adjust the relating
phase $\varphi$. So in our setup, when the boolean function $f(x)$ is a constant function,
the detector D$_2$ will be clicked, and no photons will arrive at D$_1$.
Fig. 4(a) shows our experimental results of $U_f=I$, and
Fig. 4(b) shows the results of $U_f=NOT$. Because
there is no interference in these processes, the counts of D$_1$ and D$_2$
do not change while modulating the voltage of PZT.

In the balanced-function case, $f(x)=x$ or $f(x)=inv(x)$.
We need to place the PBS$_1$ into the optical route,
so the $V$ photons and $H$ photons will travel
through the DP in different directions.
As we have discussed, when we set the
DP at $45^{\circ }$ ($-45^{\circ }$), this will be the
CNOT (Z-CNOT) gate for the input state $\left\vert \psi _{1}\right\rangle$ shown in Eq. (5).
Using the corresponding relations of Eq. (3), the theoretical analysis
of $U_f=CNOT$ is shown below.
\begin{eqnarray}
&&\left\vert \psi _{1}\right\rangle \overset{CNOT}{\longrightarrow }%
\left\vert \psi _{2}\right\rangle=(\left\vert V\right\rangle\left\vert u\right\rangle
+e^{i\varphi }\left\vert V\right\rangle\left\vert d\right\rangle +\left\vert H\right\rangle\left\vert d\right\rangle
+e^{i\varphi }\left\vert H\right\rangle\left\vert u\right\rangle)/2 \notag\\
&&\text{ \ \ \ \ }\overset{HWP_3}{\longrightarrow }\left\vert \psi _{3}\right\rangle=
[(1+e^{i\varphi })\left\vert V\right\rangle (\left\vert u\right\rangle+\left\vert d\right\rangle)
+(1-e^{i\varphi })\left\vert H\right\rangle (\left\vert u\right\rangle-\left\vert d\right\rangle)]/2\sqrt{2}.
\end{eqnarray}%
For the Z-CONT operation, we set the DP at $-45^{\circ }$. The output state is
\begin{eqnarray}
\left\vert \psi _{3}\right\rangle=
[(1+e^{i\varphi })\left\vert V\right\rangle (\left\vert u\right\rangle+\left\vert d\right\rangle)
-(1-e^{i\varphi })\left\vert H\right\rangle (\left\vert u\right\rangle-\left\vert d\right\rangle)]/2\sqrt{2}.
\end{eqnarray}%
For these two operations, we still detect the polarization qubits. Then we can get
two curves which show the photon counts of two detectors changing with the relative
phase between two spatial modes. Fig 4(c) corresponds to the CNOT operation, and Fig. 4(d) corresponds to the Z-CNOT
operation. From Eq. (8) and Eq. (9), we know that the theoretical results are
sinusoidal functions, and our experimental data fit them well.

\begin{figure}[tbh]
\includegraphics[width=6cm]{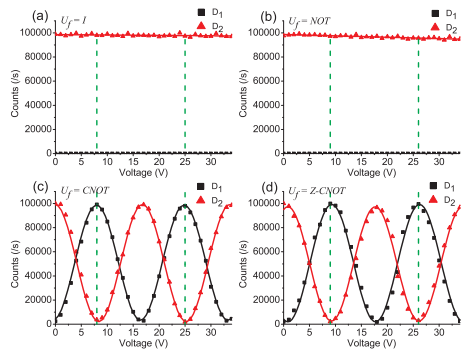}
\caption{(Color online) Experimental data of Deutsch's algorithm. Black squares
show the photon counts of D$_1$, and red triangles show the photon counts of D$_2$.
Fitting lines are also shown.
By modulating the voltage of PZT from 0 to 34 V and
every 1 V as a step, we record the photon counts of D$_1$ and D$_2$ simultaneously.
(a) Identity operation of $U_f$ for constant function $f(x)=0$. (b)
NOT operation of $U_f$ for constant function $f(x)=1$. (c) CNOT operation of $U_f$ for balanced function $f(x)=x$.
(d) Z-CNOT operation of $U_f$ for balanced function $f(x)=inv(x)$. Green dashed vertical lines are used to mark the
proper points (phases) of the initial states, which can be used to
perfectly discriminate the two kinds of functions.}
\end{figure}

Our experimental results are shown in Fig. 4. In our experiment, we make the relative phase $\varphi$ adjustable by using a
PZT controller, so the output state $\left\vert \psi _{3}\right\rangle$ contains the phase parameter $\varphi$.
When using a PBS for the projective detection, the detectors of D$_1$ and D$_2$
detect photons of different polarization: $H$ on D$_1$ and $V$ on D$_2$.
From the Eq. (8) and Eq. (9), we can see that the photon counts of D$_1$ and D$_2$ will sinusoidally vary
with the $\varphi$ being continuously changed. We set the phase range for two
periods (the voltage of PZT is adjusted from 0 to 34 V) and plot the counts-voltage curves.
From the description of Deutsch's algorithm, the input state is a certain state with certain a phase [Eq. 1].
However, we can get this state simply by setting the phase
$\varphi=(2N+1)\pi$ (adjust the PZT in proper voltages), where $N$ is an integer.
Then Eq. (8) and Eq. (9) are changed into $\left\vert \psi _{3}\right\rangle_{\pi}=\pm\left\vert H\right\rangle
(\left\vert u\right\rangle-\left\vert d\right\rangle)/\sqrt{2}$,
where `$+$' is for the CNOT operation and `$-$' is for the Z-CNOT operation.
And if we also set $\varphi=(2N+1)\pi$ in Eq. (6) and Eq. (7), we get
$\left\vert \psi _{3}\right\rangle_{\pi}=\pm\left\vert V\right\rangle
(\left\vert u\right\rangle-\left\vert d\right\rangle)/\sqrt{2}$,
where `$+$' is for the NOT operation and `$-$' is for the $I$ operation.
These results are the same as those for $\left\vert \Psi _{3}\right\rangle$,
described in Eq. (2). These proper points for Deutsch's algorithm are marked by green lines in Fig 4.
From these points, we can claim that it is a constant function when D$_1$ clicks and a balanced function when
D$_2$ clicks. Our data also show that we can only probabilistically discriminate the function if $\varphi\neq(2N+1)\pi$;
In particular, when $\varphi=2N\pi$, we cannot discriminate the two kinds of $f(x)$ at all.

Benefiting from the Sagnac interferometer, our experimental setup is very stable without any other
additional feedback control. This long time stability makes it possible to change the voltage 1 V as a step from 0 to 32 V.
We can define $\eta= |\frac{C_{D1}-C_{D2}}{C_{D1}+C_{D2}}|$ as a contrast ratio to describe the precision of our results,
where $C_{D1}$ and $C_{D2}$ denote to the photon counts of D$_1$ and D$_2$, respectively. Theoretically,
the contrast ratio $\eta$ is equal to 1. In our experiment,
for the constant functions, $\eta_{c}= 99.96\pm0.03\%$ in Fig. 4(a) and $\eta_{c}= 99.96\pm0.03\%$ in Fig. 4(b);
for the balanced functions, the contrast ratio $\eta$ equals the interference visibility;
in Fig. 4(c), $\eta_{b}= 95.76\pm0.07\%$; and in Fig. 4(c), $\eta_{b}= 96.13\pm0.07\%$.
From Fig. 4(a) and 4(b), we can see that the photon counts of D$_2$ fall with increasing voltage. This
phenomenon is mainly caused by the coupling of multi mode fibers used in the detection part.
We modulate the phase by changing the angle of the first mirror (changing the voltage of PZT). Although
the change of the angle is very tiny, it will also affect the coupling efficiency, becoming worse when photons pass though the setup.
Our experimental errors are mainly caused by
the imperfections of PBS and HWP, the interference visibility, and the effect of DP \cite{dp98,dp03}.
However, these errors can be reduced with improvement in the  experimental technique.

In conclusion, we have experimentally realized Deutsch's
algorithm using linear optical components. We can determine a
property of a function in one evaluation in the quantum case instead of
two in the classical case. When phase $\varphi=(2N+1)\pi$, we need only a
single photon as the input to judge the function $f(x)$: a constant
function when the photon is in $V$ polarization and a
balanced function when the photon is in $H$ polarization.
For the other input states, $\varphi\neq(2N+1)\pi$, we can only probably discriminate the function.
We implement the CNOT gate using a Sagnac interferometer in the SPTQ logic.
This experimental system is very stable and the experimental data are in excellent accordance with theoretical results.
We believe these can be used to perform more complex
entangled states or few-qubit quantum computation.

This work is supported by the Fundamental Research Funds for the Central Universities,
the National Fundamental Research Program (2010CB923102) and National Natural Science
Foundation of China (Grant No. 11004158, 10774139, 11074198, and 60778021).

\end{document}